\documentclass[aps,eqsecnum]{revtex4}
\usepackage{epsfig}
\newcommand{\ve}[1]{\hbox{\bf #1}}
\newcommand{\be}{\begin{equation}}
\newcommand{\ee}{\end{equation}}
\begin{document}
\title{\bf Self-gravitating fluid dynamics, instabilities and solitons}
\author{\bf  B. Semelin$^{(a)}$,  N. S\'anchez$^{(a)}$, H. J. de
Vega$^{(b)}$ \bigskip}
\affiliation
{(a) Observatoire de Paris,  Demirm, 61, Avenue de l'Observatoire,
75014 Paris,  FRANCE.
Laboratoire Associ\'e au CNRS UA 336, Observatoire de Paris et
\'Ecole Normale Sup\'erieure.  \\ (b)  Laboratoire de Physique
Th\'eorique et Hautes Energies, 
Universit\'e Paris VI, Tour 16, 1er \'etage, 4, Place Jussieu
75252 Paris, Cedex 05, FRANCE. Laboratoire Associ\'e au CNRS UMR 7589}
\date{\today} %\maketitle
\begin{abstract}
This work studies the hydrodynamics of  self-gravitating
compressible isothermal fluids. We show that the hydrodynamic
evolution equations are scale covariant in absence of viscosity. Then, we
study the evolution of the time-dependent fluctuations around singular
and regular isothermal spheres. We linearize the fluid equations
around such stationary solutions and develop a method based on the
Laplace transform to analyze their dynamical stability.   
We find that the system is stable below a critical size ($ X\sim9.0 $
in dimensionless variables) and unstable above; this criterion is the same
as the one found for the thermodynamic stability in the
canonical ensemble and it is associated to a center-to-border density ratio of 
$ 32.1 $. We prove that the value of this critical size is independent of the
Reynolds number of the system. Furthermore, we give a detailed
description of the series of successive dynamic instabilities that appear at
larger and larger sizes following the geometric progression $ X_n \sim
10.7^n ,\; n=1,2,\ldots $. Then, we search for exact solutions of the
hydrodynamic equations without viscosity: we provide analytic and
numerical axisymmetric soliton-type solutions. The stability of  exact
solutions corresponding to a collapsing filament is studied by computing linear
fluctuations. Radial fluctuations growing faster than the background 
are found for all sizes of the system. However, a critical size ($
X\sim4.5 $) appears, separating a weakly from a strongly unstable regime.
\end{abstract}
\maketitle
\section{ Introduction and results}

Understanding the  dynamics of self-gravitating fluids is a fundamental
issue in astrophysics. Indeed, the structure of the interstellar
medium as well as the formation and evolution of cosmological
structures are based on it. Compressible self-gravitating 
fluid mechanics plays a key role in both star formation and in the formation of
a fractal structure for the interstellar medium\cite{sas}. 

Self-gravitating systems can be studied at different levels.
The first is purely thermodynamic. It leads to the description of their
states of equilibrium and the study of their stability. This
approach has been 
followed in the case of the isothermal sphere configuration
\cite{Antonov,Lynden,Katz,Chandra,Padma} leading to the description of
the gravothermal instability. This behaviour appears when the size
of the sphere goes above a critical size, or, for a given size, when its
temperature drops below a critical temperature. An instability develops and
the system collapses to a dense core. In refs.\cite{nos,nos2,SVSC,gas} the
statistical mechanics of the self-gravitating gas has been
investigated with analytic (saddle point in the functional integral)
as well as Monte Carlo methods. In 
addition to the thermodynamics, this approach provides the exact equation of
state of the self-gravitating gas\cite{gas}.

The dynamics can be studied from the hydrodynamic equations of a
self-gravitating fluid. This is a set of non-linear partial
differential equations and exact solutions can be found only under
simplifying hypotheses.  For example, the case of one space dimension
has been studied in ref.\cite{Gozt} and the axisymmetric
case in ref.\cite{Inutsuka}.

Finally, the last approach to the study of self-gravitating systems
is numerical, it  has been widely followed\cite{Hernquist}.

In this paper we study the hydrodynamics of a self-gravitating fluid.
We show that the  hydrodynamic evolution equations for a
self-gravitating fluid are {\bf  covariant under scale transformations}
in absence of viscosity. That is, if we scale the space and time
variables by a constant factor in a solution of the fluid equations,
we get again a solution of the equations.

We consider the Navier-Stokes equation for an isothermal
self-gravitating fluid and restrict our study to potential
flows. Within these hypotheses we show that a 
Reynolds number appears as the only parameter that breaks the scale
covariance in the Navier-Stokes equation. Moreover we find that a
characteristic density enters the Reynolds number definition, implying
that a system with no characteristic density has no characteristic
Reynolds number; the flow is then scale covariant.

In section \ref{three} we present a general method
for the analysis of linear dynamic stability of stationary solutions of
the equations. This method is based on the Laplace transform in time of the
evolution equations. The functional determinant of the Laplace
transformed operator is studied as a function of $ s $ (the Laplace
transform parameter). Its zeros in the $s$-plane determine whether the
solution is stable or not. Unstable solutions yield zeroes with
positive real part while purely imaginary zeros or zeroes with
negative real part correspond to stable solutions (oscillating or damped). 
We apply this analysis to isothermal spheres where we first
specify the initial data and the boundary conditions.
%and then we give elements
%of proof of the existence and uniqueness of the solution.

In section \ref{four}, using the Laplace transform, 
we derive the analytic form of the fluctuations in the case of a singular
isothermal sphere background. Then, we show that the
boundary conditions build a discret spectrum of proper modes for the system. 
The stability of these modes is decided by the complex value of their 
pulsations in the Laplace transform plane. We then apply a general
numerical method based on finite differences 
to both the singular and the regular isothermal sphere cases and compute the 
pulsations as functions of the size of the system. We show that in both 
cases the dynamic stability depends on the size of the
system. Under a critical size $ X_{c_1} $ the system is stable, while above it 
becomes unstable. In dimensionless variables this size is $ X_{c_1} \sim 10.7
\, \delta $ in the singular case ($ \delta $ is the short
distance cutoff necessary in this case) and $ X_{c_1} \sim 9.0 $ in
the regular case. This critical size is the same  as the one found in
thermodynamic studies in the canonical ensemble (we used  
boundary conditions connected to the canonical ensemble). It 
is often expressed as a critical center-to-border density contrast whose value
is $ 32.1 $ for the regular isothermal sphere. However, we also
show how secondary instabilities  $ X_{c_2}, \ldots ,  X_{c_n} $
appear at larger sizes. These sizes follow
the geometric progression  $ 10.7^n \,\,(n=1,2,\ldots) $ in the singular
case and do so asymptotically ($ n \gg 1 $) in the regular case. The
density and velocity profiles of the two first instabilities are given.
Figs. 1-3 show these profiles and the spectrum of modes for the
regular and singular spheres.

In section \ref{five} we include the viscosity in the fluid equations. Then,
we study the influence of the Reynolds number on the stability criterion. The
result is that the critical sizes at which the instable modes appear are {\bf
independent} of the Reynolds number. The complicated evolution of the
pulsations of the modes as functions of both the system size and the
Reynolds number is plotted in figs. 4-6.

In section \ref{six} we turn to time-dependent background solutions. We
look for  axisymmetric soliton-type solutions. We define a soliton variable:
\be
y={\mu \, r \over 1 \pm \mu \, c_s \, t }\;,
\ee
where $ \mu $ is an inverse characteristic length ($ \mu^{-1} $ is the
Jeans' length), $ c_s $ the sound speed, $ r $ the polar radius and $
t $ the time. We reduce the equations to a single non-linear equation and we 
find non-trivial solutions such as,            
\be
\label{solitsolintro}
v_r= \pm c_s \left(\sqrt{C}+ {\mu\,r \over 1 \pm \mu\, c_s \, t} \right) \; , \qquad
\Phi= \ln\left[ { \sqrt{C} \over \mu^2 \, r^2 (1 \pm \mu \, c_s \, t)
} \right] \; . 
\ee
This solution describes an expanding or collapsing filament depending on the
choice for $\pm$. Other background
solutions are described. Then, explicit radial fluctuations around the
solution (\ref{solitsolintro}) are computed and their stability is
analyzed with the 
same method as for the isothermal spheres. Some of the modes appear to
grow faster than the background whatever the size of the system. 
A specific size,  $ X_{c_1} \sim 4.5 \, \delta $, marks
the appearance of faster growing fluctuations at larger sizes.
However, these are radial fluctuations
invariant along the axis of the symmetry; consequently they have an infinite
mass for all radial sizes of the system. Limitating the size of the
system along the symmetry axis should provide a stable behaviour for
small enough sizes.  

Section \ref{seven} is devoted to our conclusions and final remarks.

\section{ The Navier-Stokes equation for self-gravitating fluids}
\label{two}
The evolution of a self-gravitating, isothermal  fluid is described by 
a density field, $ \rho(\ve{r},t) $, a pressure field $ P(\ve{r},t) $, a 
gravitational potential $U(\ve{r},t)$ and a velocity field $
\ve{v}(\ve{r},t) $.
These four fields obey four equations: the matter conservation, the
Navier-Stokes equation, the Poisson equation and the equation of
state\cite{ldl}, respectively given by 
\be
{\partial \rho \over \partial t} + \nabla \cdot(\rho \, \ve{v})\,=\,0,
\ee
\be
{\partial \ve{v} \over \partial t} + (\ve{v}\cdot\nabla )\ve{v}=
-{\nabla P \over \rho} - \nabla U +{\eta \over \rho}\; \nabla^2 \ve{v}+
{1 \over \rho}\;(\zeta + {\eta \over 3})\;\nabla(\nabla\cdot \ve{v})\; ,
\ee
\be
\nabla^2 U \,=\, 4 \pi G \; \rho \;,
\ee
\be
c_s^2\,=\, {\partial P \over \partial \rho} \;. 
\ee
Here $ \zeta $ and $ \eta $ stand for the kinematical viscosity coefficients
and $ c_s $ is the speed of sound. We assume $ c_s $ to be a constant
which corresponds to the equation of state of a perfect gas.

>From now on, we restrict for simplicity to potential flows, which implies the
existence of a velocity potential $\Psi$. 

%
%  NEW
%
%This choice arises mainly for numerical reasons. Treating the general case
%of a velocity field with 3 degrees of freedom would lead to prohibitive
%computation cost in the numerical part of our approach.
%
%  END NEW
%
In order to work with dimensionless variables we define $\mu$, the
inverse of a characteristic length, as: 
\be
\mu^2={ 4 \pi G \rho_0 \over c_s^2}.
\ee
where $\rho_0$ is a characteristic density of the system.

We then introduce dimensionless variables and fields~:
$$
\ve{x}=\mu \; \ve{r} \quad , \quad \tau=\mu \; c_s \; t \quad,
$$
\be\label{rotor}
\Phi= \ln {\rho \over \rho_0}  \quad , \quad
\ve{v}={c_s \over \mu}\; \nabla_{\ve{\small x}} \Psi({\ve{x}},\tau) \quad.
\ee
Here, $ \nabla_{\ve{\small x}} $ is the derivative with respect to the
dimensionless  coordinate $ \ve{x} $.
We can take advantage of the potential  nature of the velocity field
to simplify the dissipation term in the Navier-Stokes
equation.  
%
%   CHANGEMENT
%
%Taking the divergence of the Navier-Stokes equation and eliminating the
%equation of state and the Poisson equation, the evolution equations
%can be written as 
%
%
Moreover, to be able to eliminate the field $U$, we now take the divergence
of Navier-Stokes equation. At the same time, to get an equivalent formulation,
we must take the rotational of Navier-Stokes equation. This last constraint
yields  the simple equation (\ref{eq4}) for non-zero viscosity. It disappear for
flows with zero viscosity. The system can now be described with
the variables $\Phi$ and $\Psi$ only as follows

\be
\label{eq1}
%\partial_\tau \nabla^2\Psi +(\partial_{i,k} \Psi)^2 + \partial_k \Psi
%\; \partial_k
%\nabla^2 \Psi  =  - \nabla^2 \Phi - e^{\Phi} +{1 \over
%Re }\nabla\cdot\left[e^{-\Phi} \nabla(\nabla^2 \Psi) \right]\quad , 
\partial_\tau \nabla^2\Psi + (\nabla \Psi \nabla)\nabla \Psi  =  
- \nabla^2 \Phi - e^{\Phi} +{1 \over
Re }\nabla\cdot\left[e^{-\Phi} \nabla(\nabla^2 \Psi) \right]\quad , 
\ee
\be
\label{eq2}
\partial_\tau \Phi + \nabla^2 \Psi + \nabla \Phi \cdot \nabla \Psi  =  0,
\ee
\be
\label{eq4}
\nabla \Phi \wedge \nabla ( \nabla^2 \Psi) = 0,
\ee
\be\label{eq3}
{1 \over Re}= \left({4 \over 3}\eta + \zeta\right)\; { \mu \over \rho_0
c_s}\quad . 
\ee
%Sumation over repeated indices is implied. Moreover we used the notation:
%\[
%(\partial_{i,k} f)^2\,=\, (\partial_i \partial_k f )^2.
%\]
$Re$ is the analogous of the usual Reynolds number for the
compressible self-gravitating flow. 

Now, let us see that the flow is
scale covariant in the absence of viscosity.
This can be proven as follows. Assume that $ \{\Phi({\ve{x}},\tau), \;
\Psi({\ve{x}},\tau) \}$ is a solution of
eqs.(\ref{eq1})-(\ref{eq4}). We  
then define the scale-transformed  fields as
\begin{eqnarray}\label{trafoesc}
\Phi_{\lambda}({\ve{x}},\tau) = \Phi(\lambda{\ve{x}},\lambda\tau)+ \ln
\lambda^2 \quad , \cr \cr
\Psi_{\lambda}({\ve{x}},\tau) = \lambda^{-1} \,
\Psi(\lambda{\ve{x}},\lambda\tau) \quad .
\end{eqnarray} 
Using that
\begin{eqnarray}
{\partial \Phi_{\lambda} \over \partial x_k} = \lambda{\partial
\Phi \over \partial y_k} \quad , \quad 
{\partial \Phi_{\lambda} \over \partial \tau } =\lambda{\partial
\Phi \over \partial {\tilde \tau} } \quad  ,\cr \cr
{\partial \Psi_{\lambda} \over \partial x_k} = {\partial
\Psi \over \partial y_k} \quad , \quad 
{\partial \Psi_{\lambda} \over \partial \tau } ={\partial
\Psi \over \partial {\tilde \tau} } \quad ,
\end{eqnarray} 
where $ y_k \equiv \lambda \, x_k $ and $ {\tilde \tau} \equiv \lambda
\tau $, we obtain that $ \Phi_{\lambda}({\ve{x}},\tau), \;
\Psi_{\lambda}({\ve{x}},\tau) $ is also a solution of
eqs.(\ref{eq1})-(\ref{eq2}) for $ Re = \infty $. The transformation
(\ref{trafoesc}) generalizes to the time dependent case the scale
transformation for $ \Phi({\ve{x}}) $ found in ref.\cite{nos2}
in the hydrostatic case. 

We find from eqs.(\ref{rotor}) and (\ref{trafoesc}) that the velocity
field transforms as,
$$
\ve{v}_{\lambda}({\ve{x}},\tau) = \ve{v}(\lambda{\ve{x}},\lambda\tau)  \quad .
$$
Under this transformation, all terms in eq.(\ref{eq2}) scale as $
\lambda $ and in eq.(\ref{eq1}) as $ \lambda^2 $ except for the 
last (viscous) term in eq.(\ref{eq1}) that scales as  $ \lambda
$. Therefore, for $ \lambda \to \infty $  scale invariance is recovered. This
corresponds to the long wavelength limit as one could expect. 

When the flow is a fluctuation, for example, a fluctuation around an
isothermal sphere with a characteristic central density, the  scale
invariance is broken. However, in this case we will see that it is 
recovered asymptotically at large distances.

\section{ The dynamic stability of isothermal spheres}
\label{three}
We study in this section the evolution of time dependent fluctuations
around an isothermal sphere. We linearize the fluid equations
(\ref{eq1})-(\ref{eq4}) around such exact static solutions
and apply  Laplace transform to solve them.
\subsection{Linearization}
Equation (\ref{eq1}) and (\ref{eq2}) have well-known static spherically 
symmetric solutions: the isothemal spheres. These are the solutions of 
the equations: 

\be \label{isoeq}
{d^2 \Phi_0 \over dx^2} + {2 \over x} {d \Phi_0 \over dx} + e^{\Phi_0}=0\;,
\ee
\be
\Psi_0=0 \; ,
\ee
where $x$ is the dimensionless radius $\mu r$. These solutions are
studied in \cite{Chandra,Katz,Padma,gas}.
%
%NOUVEAU
%
Stable isothermal spheres have a finite radius. Solutions of
eq.(\ref{isoeq}) reaching infinite radial distance suppose  an  infinite
total mass. Therefore, we shall consider the  fluid confined in a spherical
box of finite size.
%
%NOUVEAU
%

The thermodynamic stability of isothermal spheres has been
extensively studied, leading to the discovery of the gravo-thermal instability
(\cite{Antonov}, \cite{Lynden} and \cite{Katz}). Our aim is to
investigate
the dynamic instability of these states and compare with the thermodynamic
results \cite{gas}. To achieve the stability analysis we introduce
fluctuations around the isothermal spheres:

\be
\Phi(\ve{x},\tau)=\Phi_0+ \epsilon \; \phi(\ve{x},\tau) \; ,
\ee
\be
\Psi(\ve{x},\tau)= \epsilon \;\psi(\ve{x},\tau) \;  ,
\ee
where $ \epsilon $ is a small parameter.

We can then write the linearized system where only first order contributions
in $\epsilon$ are kept:

\be
\partial_{\tau} \nabla^2 \psi(\ve{x},\tau) \,=\, -\nabla^2\phi(\ve{x},\tau)
-e^{\Phi_0(x)}\;  \phi(\ve{x},\tau) +  
{1 \over Re} \; \nabla\cdot[e^{-\Phi_0(x)} \; \nabla(\nabla^2
\psi(\ve{x},\tau))] \; , 
\ee
\be \label{segunda}
\partial_{\tau} \phi(\ve{x},\tau) + \nabla^2 \psi(\ve{x},\tau) + \nabla
\Phi_0(x) \cdot  \nabla \psi(\ve{x},\tau) =0 \; ,
\ee
\be
\nabla \Phi_0 \wedge \nabla (\nabla^2 \psi) =0 \; .
\ee
Since we study fluctuations around a spherically symmetric solution,
it is natural to expand $ \psi(\ve{x},\tau) $ and $ \phi(\ve{x},\tau)
$ in spherical harmonics. 
%leads decouple the S-wave (radial) mode from the rest. 
Let us write first,  
\be
\phi(\ve{x},\tau)=\sum_{l,m} \phi_{l,m}(x,\tau) \;
Y_{l,m}(\theta,\varphi) \;, \nonumber 
\ee
\be
\psi(\ve{x},\tau)=\sum_{l,m} \psi_{l,m}(x,\tau) \;
Y_{l,m}(\theta,\varphi) \;. \nonumber 
\ee
The action of the laplacian is on $\phi_{l,m}$ and $\psi_{l,m}$ is:

\be
\nabla^2= {d^2 \over dx^2} +{2 \over x }{d \over dx} - { l(l+1) \over x^2}
\; . \nonumber 
\ee
Then, the system of equations can be written as :

\begin{eqnarray}\label{eclin}
&&\nabla^2 \partial_{\tau} \psi_l(x,\tau) + [ \nabla^2 + e^{\Phi_0} ] 
\phi_l(x,\tau) - {1 \over Re} \; \nabla\cdot[e^{-\Phi_0(x)} \; \nabla(\nabla^2
\psi_l(x,\tau))] =0 \; ,
\\ \cr
&&\partial_{\tau} \phi_l(x,\tau) + [ \nabla^2 + (\partial_x \Phi_0)
\;\partial_x]  \psi_l(x,\tau) =0 \; ,
\cr \cr\label{vincul}
&& \sum_{l,m} \nabla^2 \psi_l \,\, \partial_{\theta}Y_{l,m}  = 0\; ,  \qquad 
\sum_{l,m} \nabla^2 \psi_l \,\,\partial_{\varphi} Y_{l,m}= 0 \; . 
\end{eqnarray}

% NOUVEAU

We can see from the last condition that all $m \neq 0$
modes must be zero. This justifies labeling the radial components of $\phi$ and
$\psi$ with $l$ only. Such a cancellation of $m \neq 0$ modes happens only
in the linear approximation. Moreover, the two last conditions
are trivial for $ l=0 $; the S-wave mode is indeed decoupled from the others.
This is not the case however for $ l > 0 $ modes, which are coupled
through the condition (\ref{vincul}), if the viscosity is not zero. 
Consequently, {\bf in the viscous case}, we will
restrict ourselves to the study of S-wave  perturbations.
Since the two last conditions are automatically satisfied within
the extend of our study, we will not mention them again. 

Analytic solutions will be given in specific cases. 
We analyze below the system of equations (\ref{eclin}) by using
Laplace transformation in time, showing that this is a powerful
approach for a stability analysis.  

\subsection{ Laplace transform analysis}
\label{Laptrans}
%
% NOUVEAU
%
We are interested on fields which grow slower than some
exponential of time and therefore admit a Laplace transform for some
finite $ s > 0 $. This space of solutions is rich enough for our
analysis. 
%
% NOUVEAU
%
The Laplace transforms of our fields are  defined by:
\begin{eqnarray}\label{traflap}
\hat{\phi_l}(x,s) \,&=&\, \int_0^{+\infty} \!\!\! d\tau\; e^{-s\tau}\;
\phi_l(x,\tau) \; ,\cr \cr
\hat{\psi_l}(x,s) \,&=&\, \int_0^{+\infty} \!\!\! d\tau\; e^{-s\tau}\;
\psi_l(x,\tau) \; . 
\end{eqnarray}
The Laplace transform of the evolution equations (\ref{eclin}) takes
then the form
{\small
\be \label{galsyst} 
{\cal M}(x,s)
\left( \begin{array}{c} \hat{\psi}_l(x,s) \\ \\ \hat{\phi}_l(x,s) \end{array} 
\right) \,=\,
\left( \begin{array}{c} \nabla^2 \psi_l(x,\tau=0) \\ \\ \phi_l(x,\tau=0) 
\end{array} \right)
\ee }
where
\be
{\cal M}(x,s) = \left[ \begin{array}{cc}
s\; \nabla^2 - {1 \over Re} \; 
\nabla\cdot(e^{-\Phi_0(x)} \; \nabla\nabla^2)&  
\quad \nabla^2 + e^{\Phi_0(x)} \\ &\\
\nabla^2 + \partial_x \Phi_0\;  \partial_x & s \\
\end{array} \right]
\ee

Indeed, the inverse Laplace transforms can be written~:
\begin{eqnarray}\label{invlap}
\phi_l(x,\tau) \,&=&\, \int_{\Gamma} ds\, {e^{s\tau}\over 2 \pi i} \;
\hat{\phi_l}(x,s) \; ,\cr \cr
\psi_l(x,\tau) \,&=&\, \int_{\Gamma}  ds\, {e^{s\tau}\over 2 \pi i}\;
\hat{\psi_l}(x,s) \; . 
\end{eqnarray}
where the contour $ \Gamma $ runs upwards parallel and to the right of
the imaginary $ s $ axis. The functions $ \hat{\phi_l}(x,s) $ and $
\hat{\psi}_l(x,s) $  in the integrand of eq.(\ref{invlap}) are given
by the inverse operator $ {\cal M}(.,s)^{-1} $ acting on the r.h.s. of
eq.(\ref{galsyst}). Hence, the poles of the integrand are given by the
zeros $ s_k , \; k=1,2,3, \ldots $ of the determinant of the operator $
{\cal M}(.,s) $.

Closing the contour  $ \Gamma $
%
% NOUVEAU
%
with a path on the Re$s < 0 $ half-plane and computing the
integral in eqs.(\ref{invlap}) by residua we get for the fields a sum of terms
associated to the poles of the integrand. Then, we can push the
contour in the Re$s < 0 $ half-plane towards Re$s \to -\infty $ where
its contribution vanishes.
%
% NOUVEAU
%

Therefore, the inverse Laplace transforms
(\ref{invlap}) become sum of terms depending on time through 
$ e^{s_k \, \tau} $. If $ s_k $ has a positive
real part the mode is unstable, growing exponentially with time.
\begin{eqnarray}\label{invlap2}
\phi_l(x,\tau) \,&=& \sum_{poles} e^{s_k \, \tau} \; \mbox{res}
\hat{\phi_l}(x,s_k) \; ,\cr \cr 
\psi_l(x,\tau) \,&=&\,\sum_{poles} e^{s_k \, \tau} \; \mbox{res}
\hat{\psi_l}(x,s_k) \; . 
\end{eqnarray}
The method for analyzing the stability of the system is now clear: we  
determine the zeroes of the determinant of our operator $ {\cal M}(.,s) $
as a function of $ s $ 
for various values of $ l $ under specified boundary conditions, and we look 
for zeroes in the Re($s$)$ >0 $ half-plane of the complex $ s $-plane.  

\subsection{ Boundary conditions on an spherical domain}

Regularity at the origin requires,
\be
\label{BC}
\left. \partial_x \phi_l(x,\tau) \right|_{x=0} =0 \quad, \quad
\left. \partial_x \psi_l(x,\tau) \right|_{x=0} =0\quad .
\ee
%
%  NEW
%
As already mentioned, we confine our system in a spherical box to provide
a large distance cut-off. We consider a spherical box of radius $X$ and we
require $ \psi_l $ and the radial velocity to vanish on the surface of
the sphere. The first condition has is like a gauge condition and has
no physical effect,  while the second prevents the fluid from exiting
the confining box. 
%
%  END NEW
%
\be
\label{BCX}
\left. \partial_x \psi_l(x,\tau) \right|_{x=X} =0\quad,\quad
\psi_l(X,\tau)=0 \quad.
\ee
These conditions translate for the Laplace transforms into:
$$
\partial_x \hat{\phi}_l(0,s) = \partial_x \hat{\psi}_l(0,s) = 0
\quad, \quad \partial_x \hat{\psi}_l(X,s) =\hat{\psi}_l(X,s)=0
$$
Perturbations with non-zero total mass are in fact zero-mass perturbations 
around an isothermal sphere with a different parameter $\mu=\mu_1$. They 
{\bf must} not be included in the stability study.

We can actually prove that the mass of the field is unchanged to first
order by the perturbations. Namely, that the integral
\be
\label{masscons}
\int_0^X \phi({\vec x},\tau) \;  e^{\Phi_0(x)} \; x^2 \; dx \quad 
\ee
identically vanish. 

For perturbations with $ l \neq 0 $ the integral (\ref{masscons})
trivially vanishes upon integration over the angles. For $ l = 0 $, we
can rewrite the Laplace transform of eq.(\ref{segunda}) as follows
$$
s \; \hat{\phi}_0(x,s) + e^{-\Phi_0(x)} \nabla \left[  e^{\Phi_0(x)}
\nabla\hat{\psi}_0(x,s) \right] = 0
$$
Hence, integrating over the sphere,
$$
\left.s\int_0^X  \hat{\phi}_0(x,s) \;  e^{\Phi_0(x)} \; x^2 \; dx = -
e^{\Phi_0(x)}\; {d \hat{\psi}_0(x,s) \over dx} \right|^{x=X}_{x=0} = 0
$$
where we used eqs.(\ref{BC})-(\ref{BCX}).

In order to study the solutions of eqs.(\ref{galsyst}) with the
boundary conditions 
eq.(\ref{BC})-(\ref{BCX}), let us first reformulate eq.(\ref{galsyst}). We will
restrict this study to $ l=0 $. 

Introducing $ f(x,s)={ d{\hat{\psi}}_0 \over dx} $  we can reduce
eq. (\ref{galsyst}) to:
\begin{eqnarray} \label{eqvel}
&&{ d \over dx} \left\{\left[1+{s \over Re}e^{-\Phi_0(x)}\right]
{ d ^2 f\over dx^2}  + 
\left[{2\over x}\left(1+{s \over Re}  
e^{-\Phi_0(x)}\right)+ {d \Phi_0\over dx}\right] { d f \over dx}\right.
\cr \cr
&-&\left.\left[ s^2+ 
{2 \over x^2}\left(1+{s \over Re}e^{-\Phi_0(x)}\right)
+{2 \over x}\,{ d  \Phi_0\over dx} \right] f(x,s) \right\}= -s \;
{d^2 \psi_0 \over dx^2}(x,\tau=0)\,   \label{phidef} \cr \cr \cr
&&\hat{\phi}_0(x) =-{1 \over s}\left[{d f \over dx}(x,s)+{d \Phi_0 \over dx}
f(x,s)\right] \quad.
\end{eqnarray}
We can integrate the total derivative and drop the integration
constant thanks to eqs.(\ref{BC})-(\ref{BCX}). We find, 
\begin{eqnarray}\label{eqseco}
&&\left[1+{s \over Re}e^{-\Phi_0(x)}\right]
{ d ^2 f\over dx^2}  + 
\left[{2\over x}\left(1+{s \over Re}  
e^{-\Phi_0(x)}\right)+ {d \Phi_0\over dx}\right] { d f \over dx}\cr \cr
&&-\left[s^2+
{2 \over x^2}\left(1+{s \over Re}e^{-\Phi_0(x)}\right)
+  {2 \over x}\,{ d  \Phi_0\over dx}   \right] f(x,s)+ s \;{d \psi_0
\over dx}(x,\tau=0)  =0  \quad . 
\end{eqnarray}
The boundary conditions for $ f(x,s) $ follow from
eqs.(\ref{BC})-(\ref{BCX}): 
\be\label{condf}
f(0,s) = f(X,s) = 0 \; .
\ee
Let us first analyze this linear problem in the zero viscosity case,
namely $ Re = \infty $. Eq.(\ref{eqseco}) thus becomes,
\be\label{inhomo}
{ d ^2 f\over dx^2}  + \left[{2\over x}+ {d \Phi_0\over dx}\right] { d
f \over dx} - \left[ s^2+{2 \over x^2}+{2 \over x}\,{ d  \Phi_0\over
dx} \right] f(x,s)+ s \;{d \psi_0 \over dx}(x,\tau=0)  = 0
\ee
This ordinary second order differential equation takes a
Schr\"odinger-type form upon the transformation
$$
f(x,s) = { e^{-\frac12 \, \Phi_0(x)} \over x} \, w(x,s)\quad . 
$$
We find for the homogeneous part
\be \label{schr}
-{d^2 w \over dx^2}(x,s)+ \left[ \frac2{x^2} + \frac14\, \left({d \Phi_0\over
dx}\right)^2 + {2\over x}\, {d \Phi_0\over dx} - \frac12 \, e^{\Phi_0}
\right]w(x,s)= -s^2 \; w(x,s)
\ee
Here,
$$
V(x) \equiv \frac2{x^2} + \frac14\, \left({d \Phi_0\over
dx}\right)^2 + {2\over x}\, {d \Phi_0\over dx} - \frac12 \,
e^{\Phi_0}= \frac12\,\left({d \Phi_0\over dx} + {2 \over x} \right)^2
-\frac14\, \left({d \Phi_0\over dx}\right)^2 - \frac12 \, e^{\Phi_0}  
$$
plays the role of the `potential' and $  -s^2 $ the role of the
`energy' eigenvalue in the Schr\"odinger-type equation. 

Notice that the potential is singular and repulsive at short distances
since
$$
V(x) \buildrel{x \to 0}\over= { 2 \over x^2} + {\cal O}(1)\quad , 
$$
while it is attractive for  distances where $ {d \Phi_0\over dx} \sim
-2/x $. 
%
% LAST VERSION
%
In the case of a potential attractive at short distances, or in the case of 
an infinite domain, a continuous part could appear in the spectrum. In our
case (finite domain and repulsive potential), the spectrum will be 
discrete (see ref.\cite{ch} p 445 for a mathematical study). Notice however
that the numerical methods used in the sequel can handle both continuous
and discrete sprectrum.
%
% LAST VERSION
%

Thus, imposing the boundary condition $ w(0,s) = w(X,s) = 0 $ yields a
discrete spectrum of eigenfunctions $ w_n(x) $ with  
negative eigenvalues $ s_n^2 $ plus possibly some eigenvalues with
positive  $ s_n^2 $ corresponding to the `bound states' in eq.(\ref{schr}). 
This ensemble of eigenfunctions form a complete set.

We can now solve the inhomogeneous equation (\ref{inhomo}) by
expanding in the eigenfunctions 
$$
f_n(x) =  { e^{-\frac12 \, \Phi_0(x)} \over x} \, w_n(x)\; ,
$$
of the homogeneous problem (\ref{schr}) with boundary conditions
(\ref{condf}) and normalized as
$$
\int_0^X  x^2 \; e^{\Phi_0(x)} \; f_n(x) \; f_k(x) \; dx = \delta_{n
k} \; . 
$$
We thus write
\be \label{desarr}
f(x,s) = \sum_n c_n(s) \; f_n(x)
\ee
where the $ c_n(s) $ are for the moment arbitrary. Inserting
eq.(\ref{desarr}) into eq.(\ref{inhomo}), multiplying by $ f_k(x) $
and integrating over $ x $ yields,
$$
f(x,s) = s \sum_n \int_0^X { x'^2 \; e^{\Phi_0(x')} \; dx' \over s^2 -
s^2_n} \; f_n(x)\;f_n(x')\; {d \psi_0\over dx'}(x',\tau=0) \; .
$$
Thus, $f(x,s)$ exhibits poles in $ s^2 $ at each eigenvalue  $ s_n^2
$. Positive  $ s_n^2 $ describe instabilities as discussed at the end
of sec. IIIC since such poles imply exponentially growing fluctuations
as $ e^{|s_n| \tau } $. Negative eigenvalues $ s_n^2 = - k_n^2 $ yield
oscillating behaviour $ e^{i k_n \tau} $.

\bigskip

In the viscous case eq.(\ref{eqseco}) can be reduced to a `Schr\"odinger type'
equation with an energy dependent potential upon the transformation
$$
f(x,s) = { e^{-\frac12 \, \Phi_0(x)} \over x \; \sqrt{\alpha_s(x)}} \,
w(x,s)\quad .  
$$
where
\be
\alpha_s(x) = 1+{s \over Re}e^{-\Phi_0(x)}
\ee
We then find for the homogeneous part,
$$
-{d^2 w \over dx^2}(x,s)+ \left[ \frac2{x^2} + \frac{\left({d \Phi_0\over
dx}\right)^2}{4\; \alpha_s^2(x)} \;  + {2\over x}\, {d \Phi_0\over dx} -
\frac{e^{\Phi_0(x)}}{2 \; \alpha_s(x)} + {s \over Re} { \left({d \Phi_0\over
dx}\right)^2 e^{-\Phi_0(x)} \over 2 \; \alpha_s^2(x)}
\right]w(x,s)= -{ s^2 \over \alpha_s^2(x)}\; w(x,s)
$$
A more mathematical study of the fluctuation equations can be done
here in the spirit of ref.\cite{beyer}.

\section{The non-viscous isothermal spheres}
\label{four}

\subsection{Analytical study}
The singular isothermal sphere is an analytic solution to eq
(\ref{isoeq}) given by,
\be
\Phi_0= \ln  { 2 \over x^2 }  \; .
\ee
This solution has an infinite total mass and produces infinite
pressure at the center. To avoid this divergence we introduce a short distance 
cutoff $\delta$. The spherical box provides the long distance cutoff.
%
% Modif
%
We study the singular sphere background as an analytic model for the
regular sphere. In this spirit we rewrite the boundary conditions  as:

\begin{eqnarray}
\label{BC1}
&&\left. \partial_x \hat{\psi}_l(x,s) \right|_{x=\delta} \; =0\quad  ,\quad 
\left. \partial_x \hat{\phi}_l(x,s) \right|_{x=\delta} =0\; ,
\cr \cr
&&\left. \partial_x \hat{\psi}_l(x,s) \right|_{x=X} =0
\quad  ,\quad
\hat{\psi}_l(X,s)=0\; .
\end{eqnarray}

Conditions at $x=\delta$ are straightforward translations of
conditions at $x=0$. 

The advantage of this analytic background solution is that we can write down 
analytic expressions for the fluctuations. Indeed for $ l=0 $ 
eq.(\ref{schr}) reduces to:

\be
\left[ {d^2  \over d x^2} +  {2 \over x^2 } - s^2 \right] f(x,s)=0 \; ,
\ee
where $ f(x,s)= w(x,s) $ in this case, is the dimensionless radial
velocity of the $l=0$ mode. In  dimensionful variables, the radial velocity
fluctuations for $ l=0 $ are:

\begin{eqnarray}\label{sing1}
v_r(r,t)&=&c_s\; \left(A_1 e^{i \omega t} + A_2 e^{-i \omega t}\right)\;
\sqrt{kr} \;
\left[ B_1 \; J_{\nu}(kr) + B_2 \; J_{-\nu}(kr)\right] \; , \cr \cr
s &=& i k \quad , \quad\omega^2=k^2 c_s^2 \quad , \quad  \nu= {i
\sqrt{7} \over 2} \quad . 
\end{eqnarray}
where $ J_{\nu}(z) $ are Bessel functions. The corresponding
fluctuation $\phi(r,t)$ takes the form

\be\label{sing2}
\phi(r,t)=\left(A^{\prime}_1 e^{i \omega t} + A^{\prime}_2 e^{-i
\omega t}\right)\; \sqrt{kr} \;
\left[ B^{\prime}_1 \; {d J_{\nu}\over dr} (kr) 
+ B^{\prime}_2 \; {d J_{-\nu}\over dr}(kr)\right] \quad .
\ee

The boundary conditions (\ref{condf}) yield a system of two
homogeneous equations for the constants $ B_1, \; B_2 $, 
\begin{eqnarray}
&&B_1 \; J_{\nu}(kX) + B_2 \; J_{-\nu}(kX) = 0 \cr \cr
&&B_1 \; J_{\nu}(k\delta) + B_2 \; J_{-\nu}(k\delta) = 0
\end{eqnarray}
Therefore, in order to have a non-trivial solution we impose the
trascendental equation, 
\be\label{ecbes}
{ J_{\nu}(kX) \over J_{-\nu}(kX)} = {J_{\nu}(k\delta)\over
J_{-\nu}(k\delta)} 
\ee
This equation is fulfilled for
specific values of $k$, yielding a  discret spectrum for the
modes. 

Real solutions $ k_n $ of eq.(\ref{ecbes}) can be obtained
asymptotically using the Mac Mahon expansion \cite{pet}. Depending on
the value of $ X/\delta $ there can be solutions with real or with
imaginary $ k_n $. The threshold of instabilities precisely appears
when the eigenvalues go through $ k = 0 $ turning from real to
imaginary. In such limit eq.(\ref{ecbes}) yields
$$
\sin\left[ {\sqrt7 \over 2 } \log{X_c \over \delta} \right] = 0
$$
with the exact solutions
$$
{X_{c_l} \over \delta} = e^{2\pi l \over \sqrt7}= (10.749\ldots)^l
\quad , \quad l=1,2,3,\ldots 
$$
Each time $ {X \over \delta} $ crosses each of these thresholds
upwards, a new instable eigenvalue appears. The first instability
occurs for $ l = 1 $. These points exactly coincide with those found
in ref.\cite{SVSC}.

We have checked that the general method that we will 
describe in section (\ref{metnum}) gives numerically  the same results
as eqs.(\ref{sing1})-(\ref{ecbes}). 

\begin{figure}[t]
\begin{center}
\epsfig{width=17cm,file=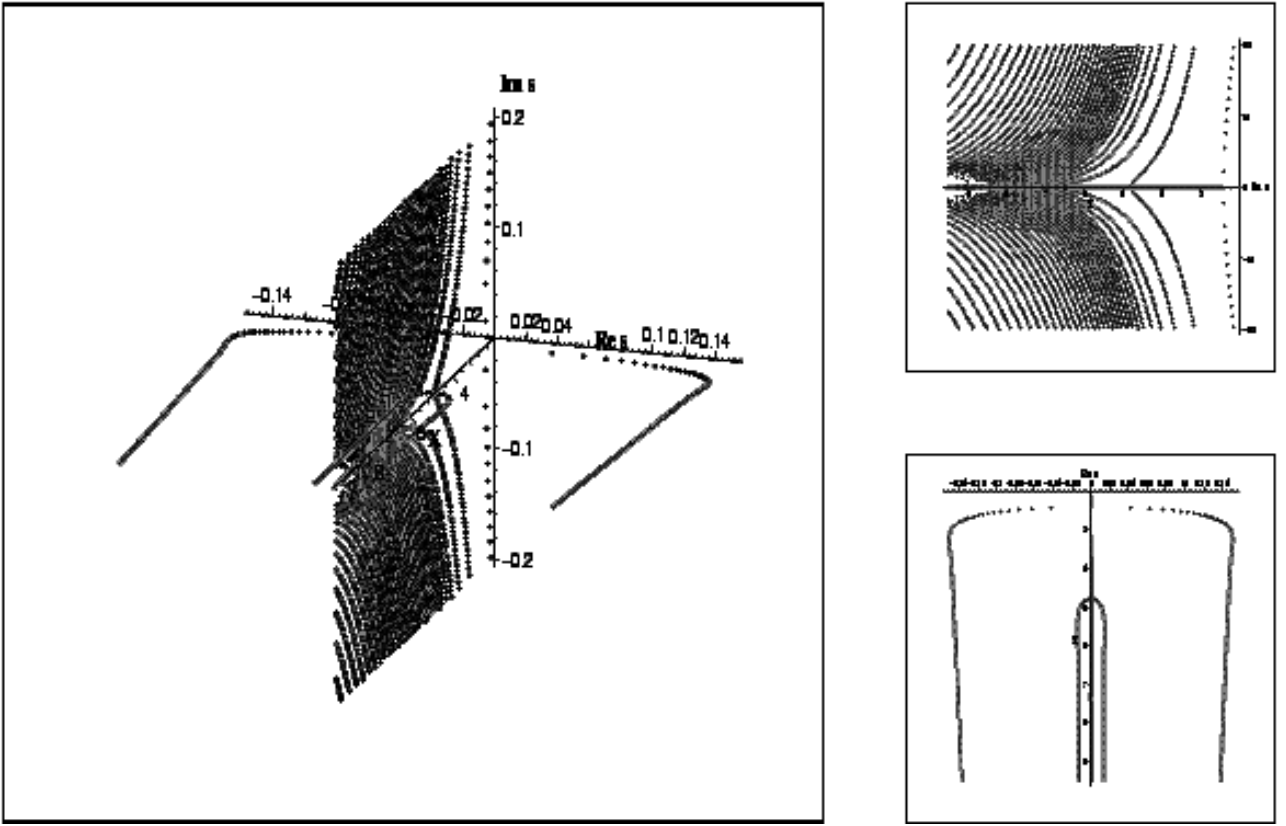}
\caption{Spectrum of the modes around the singular isothermal sphere. 
The real and 
imaginary parts of the pulsations of the modes are plotted in the 3d-view for
a range of values of the size $X$. All scales are {\bf logarithmic} (
$\ln(1+x)$ actually). Views from the right and from above are also plotted.} 
\label{absing_fig}
\end{center}
\end{figure}
\begin{figure}[t]
\begin{center}
\epsfig{width=17cm,file=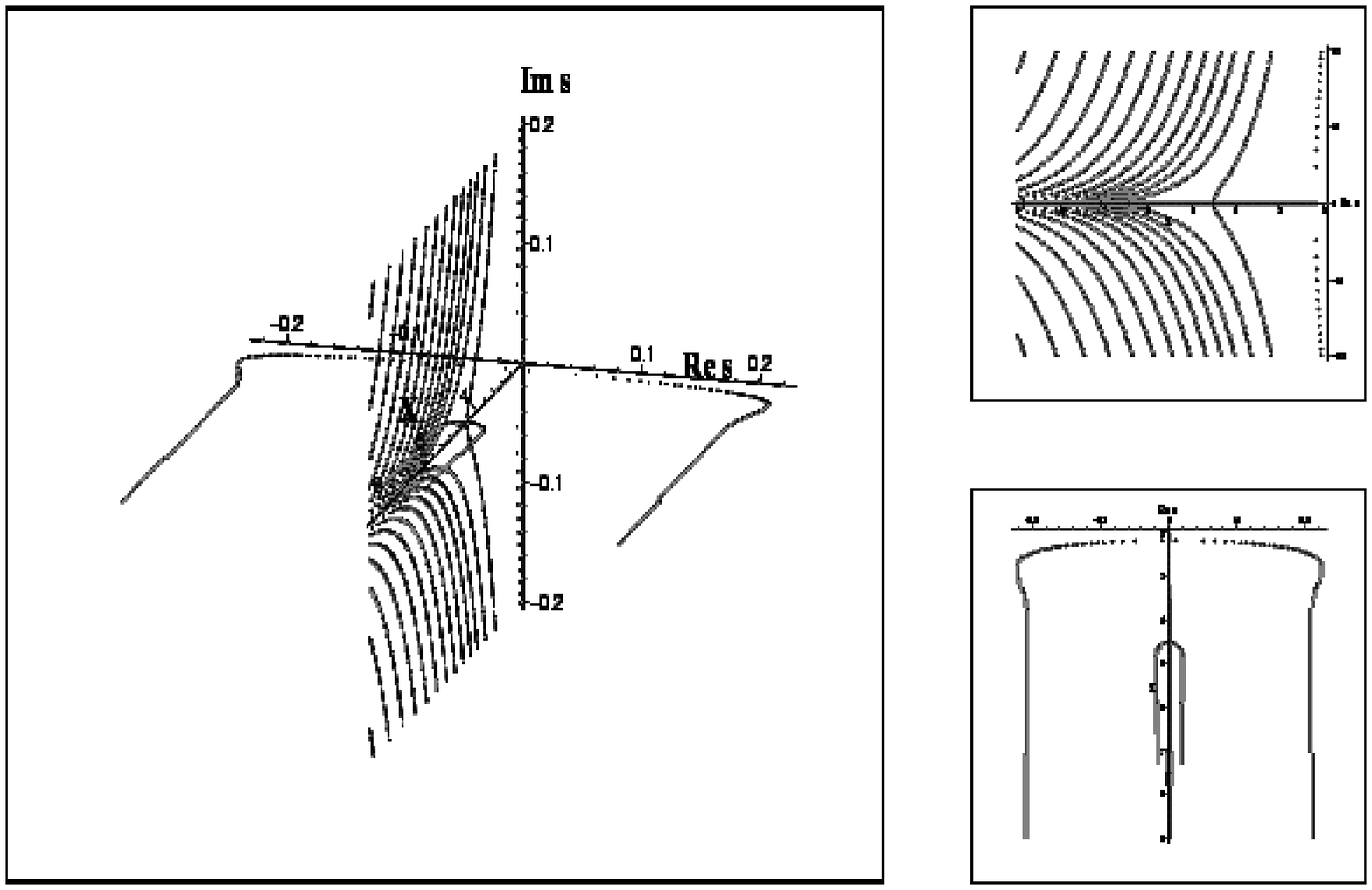}
\caption{Spectrum of the modes around the regular isothermal spheres.
Explanations are the
same as in Fig \ref{absing_fig}.}
\label{abreg_fig}
\end{center}
\end{figure}

%
% Nouvelle section
%
\subsection{Numerical method}
\label{metnum}
The general numerical method, consists  in replacing the derivatives
by finite differences turning eq.(\ref{galsyst}) and the
boundary conditions into a $ N \times N \; (N \gg 1 ) $ linear system.
More specifically, $ \hat{\phi} $ and $ \hat{\psi} $ as functions of
$ x $ are represented by their
values on a discret set of points $\{x_k,1 \le k \le N\}$.
The differential operator are
then represented by finite differences. Evaluating the differential equation
in each point produces an algebraic equation. Boundary conditions provide the
missing equations at the boundary of the domain, where the
differential operator 
cannot be evaluated. As explained in section \ref{Laptrans}, we then 
evaluate the determinant of the system in a given domain of $ X/\delta
,  \; s $ and $ Re $ and look for cancellations. 

As the parameter domain is huge (four-dimensional), we must limit the
number $ N $ in the determinant. It has been set to $~40$ in most 
computations. However, different values (up to 200) have been used to
investigated the stability of the results around randomly chosen
cancellations of 
the determinant, and more specifically in the region of the 
stability-instability transition. No ill-behaviour has been detected. 

\subsection{Stability of the singular sphere}
\label{singsph}

Cancellations of the determinant of the differential operator happen
for discret values of the parameter $ s $. If one of these eigenvalues
$ s $ has a positive real part, then the associated
mode  grows with time and the  system is unstable. It is known from
thermodynamic studies that for a given central
density, the stability depends on the size of the sphere. Thus,  we will
investigate for the zeroes of the determinant in a three-dimensional domain
$ (s,X) \in C \times R  $. The zeroes are shown on % \mathbb
fig.\ref{absing_fig} for $ l=0 $.

The first result of fig.\ref{absing_fig} is that it shows a symmetry around 
the Re($s$)$=0$ axis; we have a 
set of coupled modes with opposite pulsations. Then, we have checked that 
the pulsations of the spectrum of modes have purely imaginary values
for $ X/\delta $ smaller than a critical value $ X_{c_1}/\delta $.
This range of $ X/\delta $ values in mostly outside the
domain pictured on fig.\ref{absing_fig} since the behaviour of the system
is simple there (however $ \ln(X_{c_1}/\delta+1)\sim 2.4 $ is visible on the
figures). In this regime the modes are oscillating stationary modes
and the system is stable. 
As $ X $ increases above $ X_{c_1} $, the size of the system grows and a
first mode encounters a branching point and develops a real positive 
pulsation. This means
that the mode grows exponentially and the system becomes
unstable. Then, for successive 
critical values $ X_{c_i} , \; i=1,2,\ldots $ corresponding to larger
and larger sizes of the  
system, new modes become unstable. It is interesting to mention that the
critical values appear when the rate of the larger and lower cutoffs is 
$  X_{c_n}/\delta \sim 10.7^n ,\; n=1,2,\ldots$. This geometric
progression has been found in the thermodynamic study in ref.\cite{SVSC}. 
When $X$ goes to infinity, an infinite number of unstable modes appear, which
is not surprising since the system then has an infinite mass. 
%
% NOUVEAU
%
Obviously, this limit is unphysical.
%
%NOUVEAU
% 

Non-radial perturbations have been investigated ($l \ne 0 $). These
perturbations are always stable and oscillating. 
%In these cases, the
%zero-mass constraint has no effect. However, additional constraints apply.
%For example, the case $ l=1 $ could show a  
%pseudo-instability associated with an overall (irrelevant) translation of the
%system, but such a mode can be excluded with the correct constraint.

\subsection{Stability of the regular isothermal spheres}
  
Regular solutions of eq.(\ref{isoeq}) have been widely studied in spite
of the fact
that they do not have an analytic expression. In this case it is not 
necessary to introduce a short distance cutoff in the study of the fluctuations
since the  background solution is regular at the origin. We keep
considering the solution in a large but finite sphere of radius $ X
$. Thus we will use the same boundary conditions as in the singular 
case, taking $\delta=0$. 

We again apply the general numerical method described above. We
compute  the spectrum of pulsations of the modes for varying values of
the size $ X $ of the sphere (see fig.\ref{abreg_fig}).

The first point is that the behaviour is very similar to the singular case.
The interpretation of fig.\ref{abreg_fig} is the same as that of
fig.\ref{absing_fig}. This means that the singular isothermal sphere
with short distance cutoff is a good analytic 
analogue of the regular isothermal sphere. The only difference in
the behaviour of the fluctuations is in the specific values of the
$ X_{c_i} $. In the regular case, the instability appears at a smaller size ($
X_{c_1}/\delta\sim 9. $) than in the singular case ($ X_{c_1}/\delta\sim 10.7
$).  
%The value in the singular case is given for $\delta=1$ since the
%actual parameter is the ratio $ {X \over \delta} $. 
As expected, the value $ X_{c_1}/\delta\sim 9.0 $ corresponds to a ratio of $
32.1 $ between the central and border densities of the background. This
critical ratio is the same as in the study of the thermodynamic
stability in the canonical ensemble\cite{Katz}. 
Finally the numerical study suggests that the ratio $10.7$ between
two successive $ X_{c_i} $ is asymptotically obtained at large $X$ in
the regular case, as it must be.

\subsection{ Profile of the instable modes}

Having established that the dynamic instability appears for the same
center-to-border density contrast as in the thermodynamic instability, we 
now check if the profile of the first instable mode matches the profile of the
thermodynamic instability, and we investigate the profile of other instable 
modes.

\begin{figure}[t]
\begin{center}
\epsfig{width=13cm,file=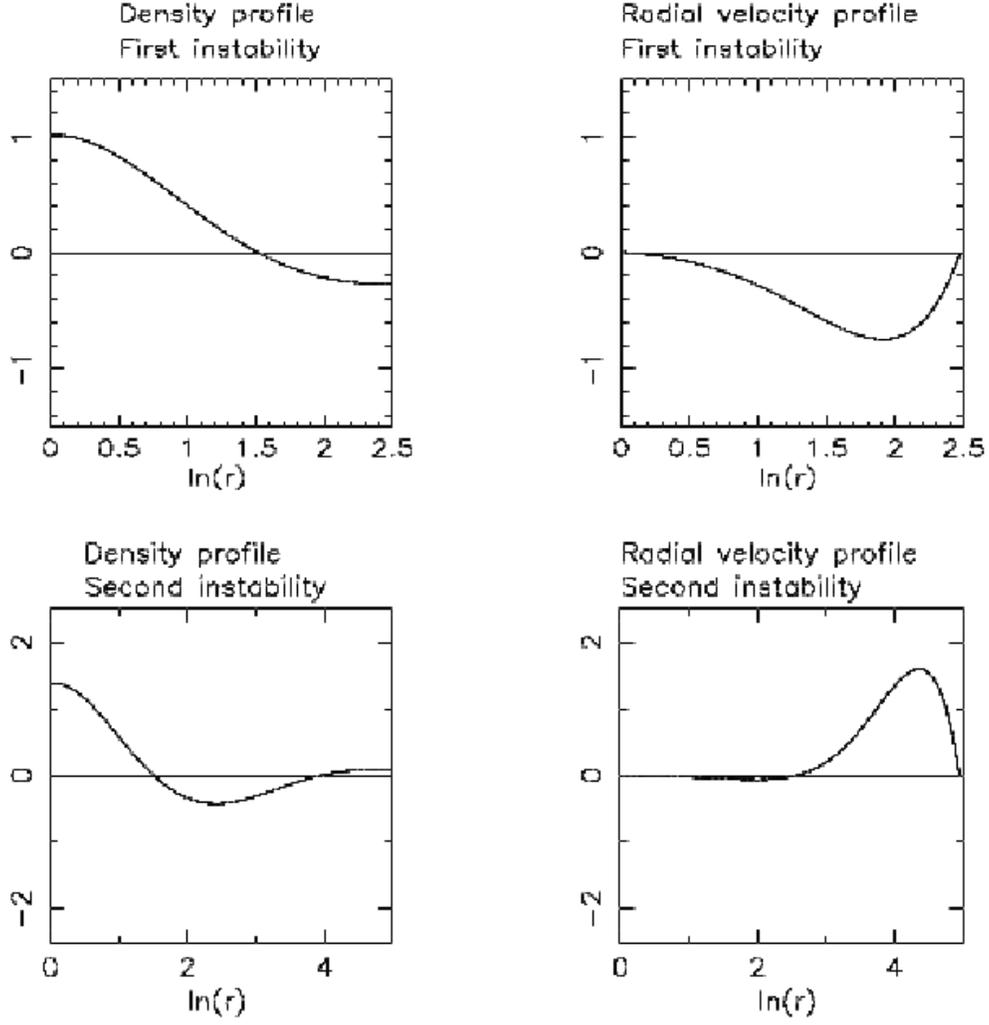}
\end{center}
\caption{Radial profiles of the first two instabilities, at sizes just above
their emergence, plotted for arbitrary amplitudes.}
\label{modes_fig}
\end{figure}

Those profiles are given, when we use the finite differences method, 
by the coordinates
of the vector of the one-dimensional kernel of our operator (boundary
conditions 
included). For a given size,
we are thus able to determine the profile of each mode associated with a
zero of the determinant for specific values of $s$.
On fig. \ref{modes_fig} the profiles are given for the first and second
instabilities for sizes just above their critical sizes. As can be
seen, the first instability is a spherical collapse forming a dense
core. Its density profile is similar to the profile of the
thermodynamic instability \cite{Padma}. The second instability
combines a core collapse and the ejection of a shell.

\section{Effect of the viscosity on the dynamics and stability criteria}
\label{five}

It is now interesting to establish how viscosity alters the dynamics of the
fluctuations. We first proceed to an analytic study of  the
fluctuations and  then perform a thorough numerical analysis.

\subsection{ Analytic study}

Here we will study the fluctuations around the singular isothermal sphere
with a non-zero viscosity. We consider only radial perturbations since
it simplifies the equations and contains the essential information
about stability. Using the variables:
\be
y=s \, x\; , \qquad \qquad f(y)=\nabla \psi (sx) ,
\ee
where $f$ is the (dimensionless) radial velocity, the system
(\ref{galsyst}) becomes

\be
\left(1+{1 \over s \; Re} {y^2 \over 2} \right) {d^2 f \over dy^2} + {y \over s
\; Re} {df \over dy} +\left({2 \over y^2 }-1-{1 \over s Re}\right) f =0 \; .
\ee

It is interesting to notice that all the dependence on the parameters
is through the combination $ s \, Re $, which is the turbulent-like Reynolds
number associated to a mode of pulsation $s$.
The asymptotic behaviour of the solutions at large distance is:

\be
y \gg 1 \; , \; y \gg s \; Re\; :\;   f(y)=y^{ -1 \pm \sqrt{9+8\,s\,Re}
\over 2} \;. 
\ee

For $sRe > -1$, the component with the upper sign grows at large
radius while the other component decays. For negative $ sRe < -1 $, 
both solutions decay at large radius. The interval 
$ -1>s \, Re >-9/8 $ is the only range giving strictly decaying
behaviours at large 
radius. In the other cases ($ s \, Re < -9/8$ or $ s $ complex) we have
spatially oscillating and decaying behaviours.  
It is easy to check that at small distances one recovers the zero-viscosity
solutions.
This, of course, does not take the boundary conditions into account. 
We provide in the next section quantitative results from  numerical 
calculations. 

\subsection{Numerical study}

\begin{figure}[p!]
\begin{center}
\epsfig{width=17cm,file=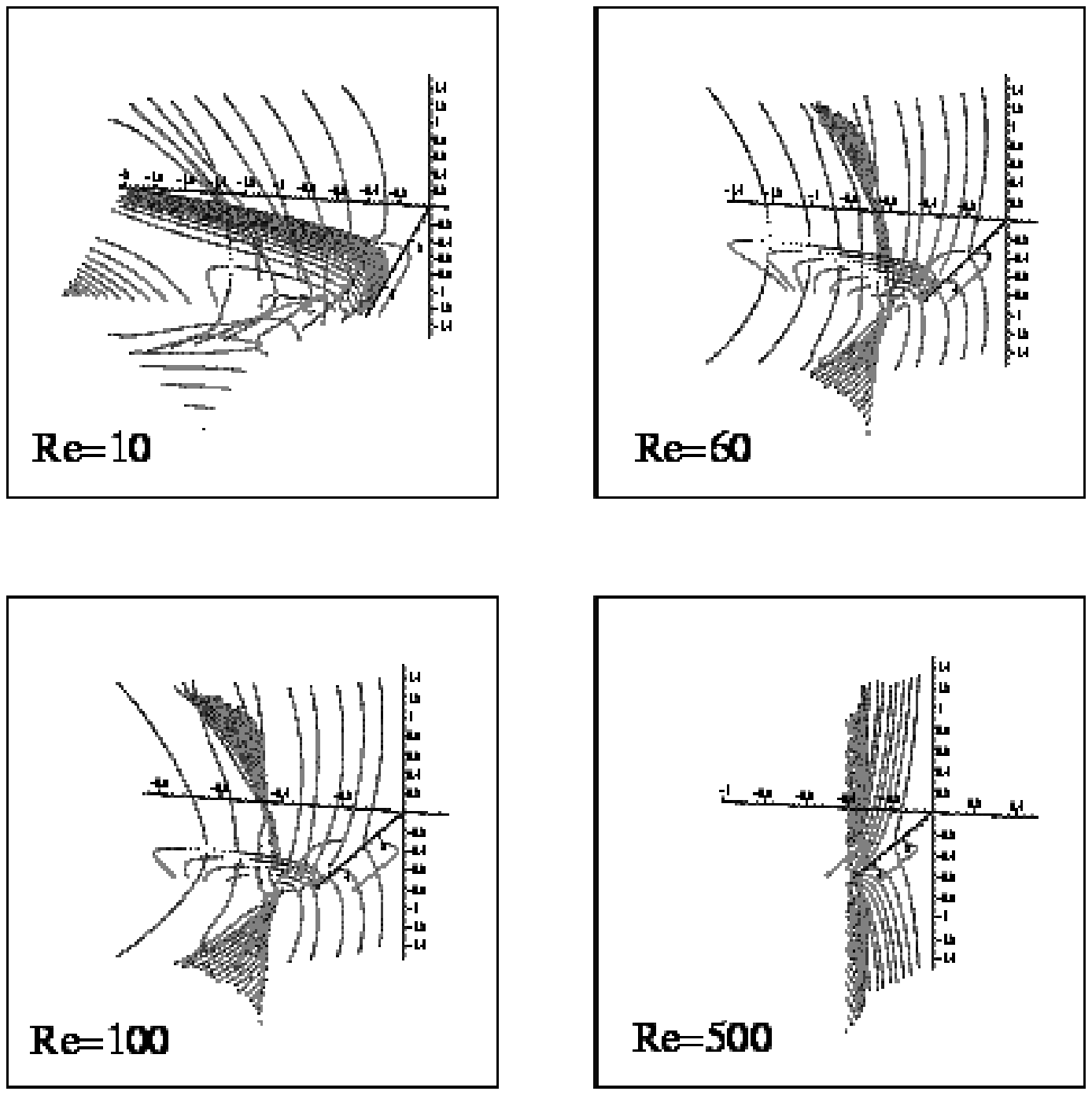}
\caption{\small Spectrum of viscous modes around the singular
isothermal sphere. 
The real and imaginary parts of the pulsations of the modes are plotted in
3d-views for a range of values of the size $X/\delta$, and four values
of the Reynolds number. All scales are logarithmic.}
\label{res3d}
\end{center}
\end{figure}

\begin{figure}[p!]
\begin{center}
\epsfig{width=17cm,file=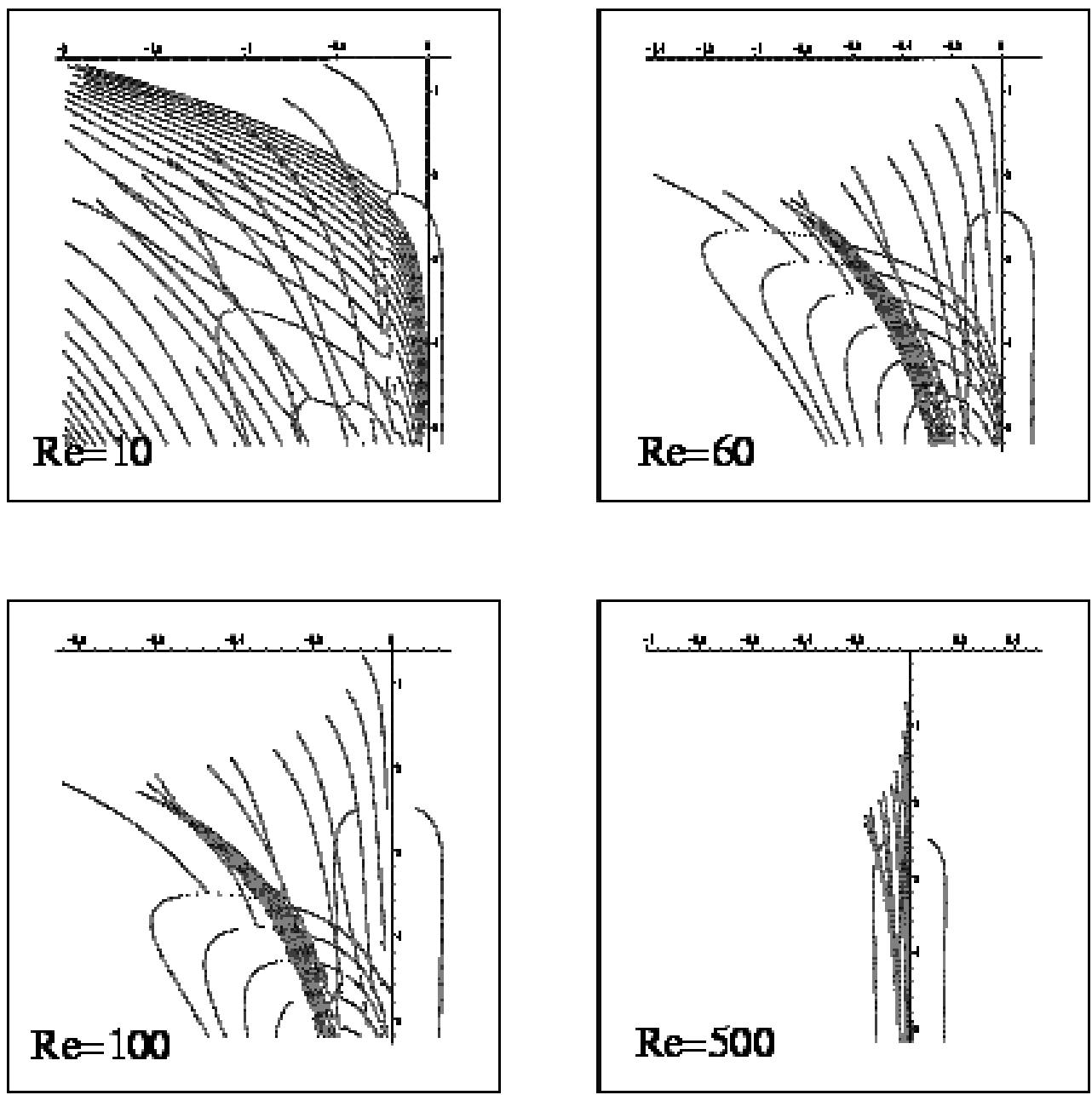}
\caption{\small Upper views of fig. \ref{res3d}. The real parts
of the pulsations
of the modes appear as functions of the ratio $X/\delta$. Positive real parts
appear for the same values of $ X/\delta\; (\ln(1+X/\delta)\sim 2.4)$,
independently of the Reynolds number.}
\label{resRsr}
\end{center}
\end{figure}

\begin{figure}[h!]
\begin{center}
\epsfig{width=17cm,file=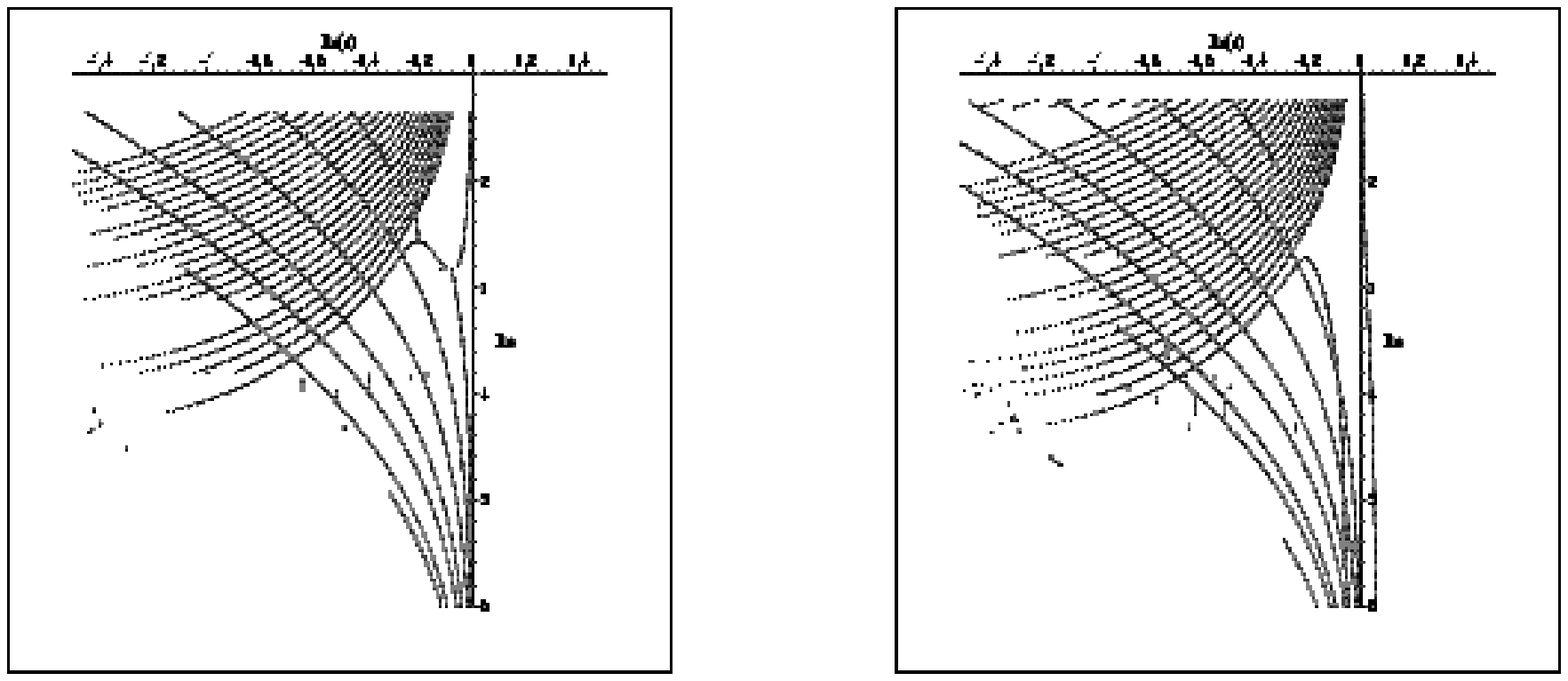}
\caption{The real parts of the pulsations of the modes are plotted as functions
of the logarithm of the Reynolds numbers for fixed sizes:  10.5 on the left
and 11.6 on the right.}
\label{fxsize}
\end{center}
\end{figure}

The introduction of the viscosity, although formally cumbersome, fits
with no particular 
difficulty in the finite difference scheme. 
%
% Nouveau commentaire
%
In particular, it does not
change the actual order of the system of differential equations as can be
seen from eq.(\ref{eqvel}).
The spectrum of modes for a
continuous range of the larger cutoff $X$ can be computed for different values
of the Reynolds number. These results are shown on fig.\ref{res3d} and
fig.\ref{resRsr}.  We show
in fig.\ref{fxsize} the evolution of the spectrum at fixed sizes, for
a continuous range of values of the Reynolds number. 

Obviously, the diagrams are more complex in the viscous case than in the
ideal case, especially as we go to Reynolds numbers close to one. 
Reynolds numbers as small as 0.1 have been investigated but are not
plotted since 
they show no specificity and are unlikely values for a self-gravitating gas. 
We have three main kind of modes: oscillating-decaying modes,
decaying modes, and growing modes. However, all the values of the critical
sizes associated with the onset of instabilities  are {\bf independent} of the 
Reynolds number. This can be guessed on fig.\ref{resRsr}, and checked on fig
\ref{fxsize} for $X_{c_1}$. The stability
of the isothermal sphere {\bf does not depend} on the viscosity.
Besides,  figs.\ref{res3d} and \ref{resRsr} clearly show that the mode
associated with the first instability is much less affected by the viscosity 
than the others. 


\newpage

The real parts of the modes are plotted on fig.\ref{fxsize} as functions of
the Reynolds number with logarithmic scales, for two different sizes: 10.5 and
11.6 . In the first diagram all modes are stable
independently  of the Reynolds number and in the second there is exactly
one unstable mode for all values of the Reynolds number. This is a confirmation
that the dynamic stability criterion depends only on the size of the system
and not on the value of the Reynolds number.

\section{ Soliton-type solutions in the self-gravitating gas}
\label{six}

The appearance of solitons in an isothermal self-gravitating gas has been 
studied in one spatial dimension in ref.\cite{Gozt}. 
In this case, the system could be reduced to a sine-Gordon
equation. Here we will study the case of axisymmetric soliton-type
solutions in $3 + 1$ dimensions.

\subsection{Soliton-type equations}
We will consider $z$-independent solutions of the evolution equations
(\ref{eq1})-(\ref{eq3}) without viscosity ($ Re = \infty $). It is
useful to make the following change of variables, 
\be
(r,t) \qquad \longrightarrow \qquad \left( y={r \over r_0(t)}\; ,\;t
\right) \; , 
\ee
where $ r_0(t) $ is a time dependent characteristic length. The simplest 
non-trivial choice for $ r_0(t) $ is a linear dependence on
time. Using the characteristic parameters of the system this leads the
soliton  variable: 

\be
\label{solitvar}
y={\mu\;  r \over 1 \pm \mu\; c_s\; t}= {x \over 1 \pm \tau} \; .
\ee
We  write the potential and radial velocity field as:
\be
\label{fields}
\Phi=\ln\left[{f(y) \over (1 \pm  \tau)^2 }\right] \; , \qquad  \qquad
{d \Psi\over dx}=h(y)\pm y \; . 
\ee
Then, the matter conservation equation (\ref{eq2}) yields
$$
{ h'(y) \over h(y)} + {1 \over y} + { f'(y) \over f(y)} =0 \quad ,
$$
which can be immeadiately integrated as
\be
\label{consmat}
f(y)\; h(y)\; y\,=\,\mp C  \; .
\ee
The symbol $\mp$ in front of $C$ is introduced for later convenience. Each
possibility is associated with one of the choices for the soliton variable.
$C$ going to zero means that the density goes to zero. Inserting
eqs.(\ref{fields}) and (\ref{consmat}) in the Euler equation of motion
(\ref{eq1})  gives the non linear  equation:
\be
\label{soliteq}
h\left(y \left(h h^{\prime} \pm h-{h^{\prime} \over h}\right)
\right)^{\prime}=\pm C \; . 
\ee
It is possible to derive some analytic solutions of this equation.

\subsection{Analytic solutions}

The simplest solution to eq.(\ref{soliteq}) is $ h(y) = \mp \sqrt{C} $. In
terms of physical fields this produces non-trivial solutions:
\be
\label{solitsol}
\Psi_0^{\prime}(r,t)= \pm c_s \left(-\sqrt{C}+ {r \over \mu^{-1} \pm
c_s t} \right)  \; , \; 
\Phi_0(r,t)= \ln\left[  {-\sqrt{C} \over \mu^2 r^2 (1 \pm \mu c_s t) }
\right] = \ln{2 \over x^2} + \ln(\sqrt{ C} y) \; .
\ee

These two solutions are filaments, one expanding and the other collapsing
in a finite time. This solution is invariant under the following scale
transformations
\be
\label{transf}
C \longrightarrow \lambda\; C\; , \qquad c_s \longrightarrow {c_s
\over \lambda} \; , \qquad \mu \longrightarrow \lambda\; \mu \; .
\ee
Hence, in the calculation of the fluctuations [sec. VI C], we can set
for example $ \sqrt{C}=1 $.

\begin{figure}[b!]
\begin{center}
\epsfig{width=17cm,file=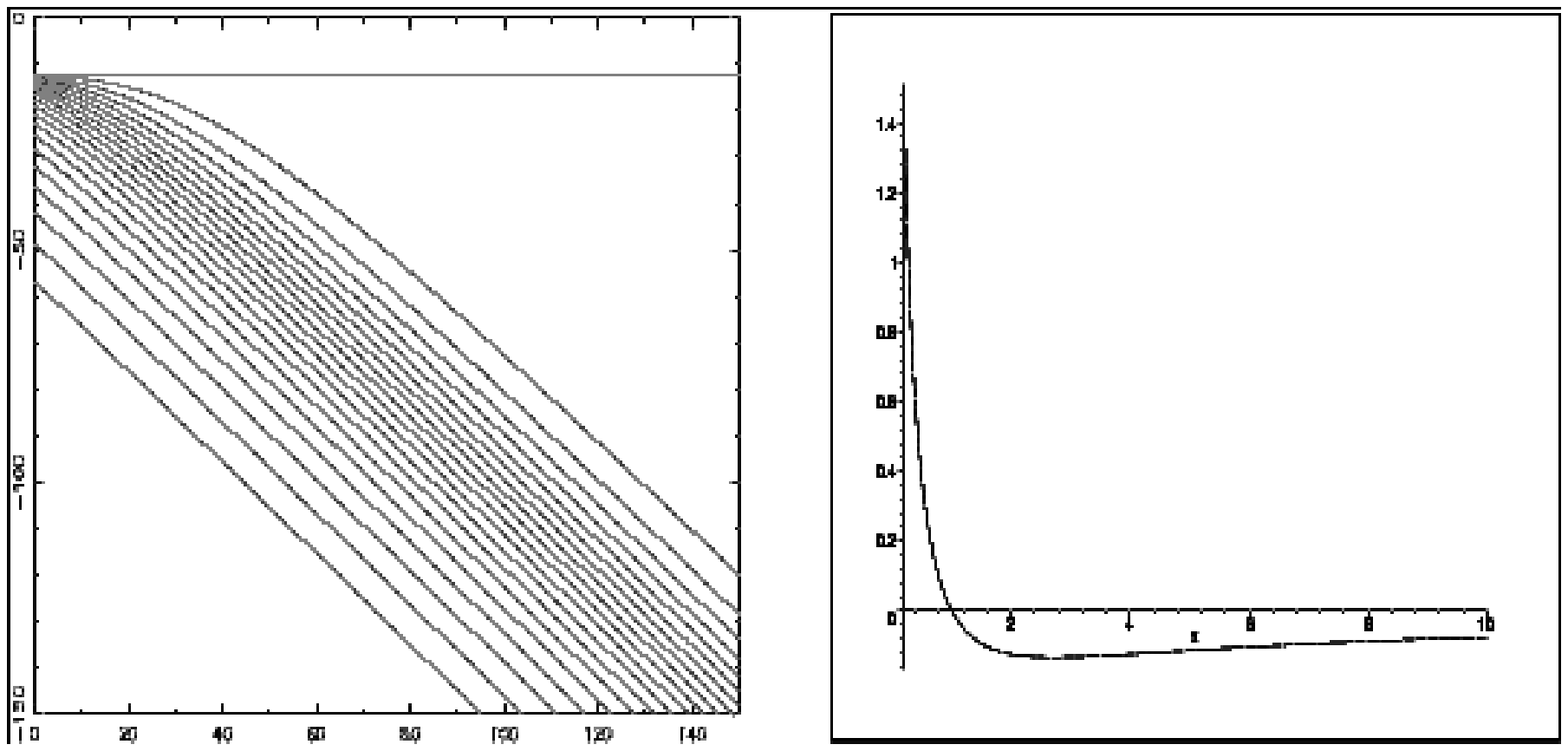}
\caption{On the left solutions of eq.(\ref{soliteq}) are plotted  for various
initial conditions. On the right the function $h_1$ is plotted for $B=1$.}
\label{solitonfig}
\end{center}
\end{figure}

Guided by numerical integration (see fig. \ref{solitonfig}), another type of 
solution can be found for low densities. It is possible to build a
perturbative solution of eq.(\ref{soliteq}) around the exact solution  $ C=0
,\;  h(y)=\mp y $. Since the system is ill-defined at $ C=0 $, we will
consider $ C \ll 1 $ and we will develop $ h(y) $ perturbatively in $ C $.
\be \label{hache}
h(y)=\mp y\left[ 1+C\; h_1(y)+{\cal O}(C^2) \,\right] \; .
\ee
Then $ h_1(y) $ obeys the simple equation:
\be
y^2\;h_1+y(y^2-1)\;h_1^{\prime}=\pm \ln {B \over y}  \; , 
\ee
which has the general solution,
\be\label{hache1}
h_1(y)={1 \over\sqrt{|y^2 -1|} }\left(A \pm \int {dy \; \ln\left({B
\over y}\right) \over y \sqrt{|y^2-1|} } \right) \; . 
\ee
We see that in general $ h_1(y) $ is singular at $ y = \pm 1 $, that is, on the
wavefront $ x = | 1 \pm \tau | $. Regular solutions can be obtained by
appropriate choices of $A$ and $B$. We plot in  fig.(\ref{solitonfig})
a regular solution $ h_1(y) $ for $ B=1 $.

We obtain from eqs.(\ref{hache}) and  (\ref{hache1}) the physical fields:
\be
\Psi=\pm c_s \; C\; y\; h_1(y)\quad  ,  \quad \Phi=\ln \left\{ {C \over x^2
}[1-C\;h_1(y)] \right\}\; . 
\ee
It is important to notice that this perturbative development fails at short
distances ($y \to 0 $) where the density is not small anymore.

\subsection{Fluctuations around a soliton-like solution}

We intend to assess the stability of the solution (\ref{solitsol}). By
stability we mean that the growth of the fluctuations should be slower
than the  growth of the background.

We compute here the explicit form of the radial fluctuations around the
soliton-like solution described by eq.(\ref{solitsol}).
The collapsing filament case is studied. We define the two 
independent variables:
\be
y= {x \over 1 - \tau} \; , \qquad z = 1 - \tau \; .
\ee
We consider the perturbative expansion
\be
\Phi(x,\tau)= \Phi_0 + \epsilon \; f(y,z)\; , 
\qquad {d \Psi \over dx}(x,\tau)={d \Psi_0 \over dx} -c_s  \;
\epsilon \;  g(y,z) \; . 
\ee
We can then linearize the evolution equations and we find
\be
\label{p1}
\sqrt{C} y {\partial^2 g \over \partial y^2}- yz {\partial^2 g \over \partial y
\partial z} - (-\sqrt{C}+y) {\partial g \over \partial y} -z {\partial g \over \partial
z} -g = y{\partial^2 f \over \partial y^2} + {\partial f \over \partial y}+
\sqrt{C} f \; ,
\ee
 
\be
\label{p2}
-\sqrt{C} {\partial f \over \partial y} + z {\partial f \over \partial z} +{\partial g
\over \partial y} =0 \; .
\ee
These equations are homogeneous in $z$ so their solutions can be written as a 
sum of solutions of the type:

\be
\label{varsep}
f(y,z)=f(y) \; z^{\alpha}\; , \qquad g(y,z)=g(y) \; z^{\alpha}\; , \qquad
\alpha \in  C\; . %\mathbb{C}
\ee
Inserting these expressions in eqs.(\ref{p1}) and (\ref{p2}) and eliminating
$g$, gives an equation for $f$:

\be
x(C-1) \; f^{\prime\prime\prime} + \left[ 2(C-1)-(2\alpha+1)\sqrt{C}x\right]
 \; f^{\prime\prime} + \left[-(4\alpha+3)\sqrt{C} + \alpha(\alpha+1)x \right] \;
f^{\prime}+2(\alpha+1)\alpha \; f=0 \; .
\ee

This equation can be integrated in the general case. However,
according to the transformation (\ref{transf}) we can choose the case
$\sqrt{C}=1$. In this case the solutions take the following form: 
\be
f(y)=B_1\; \Phi\left(\lambda_1,\nu_1,\beta y\right)
+ B_2\; y^{\gamma}\; \Phi\left(\lambda_2,\nu_2, \beta y \right)\; .
\label{solfluct}
\ee
where,
\be
\lambda_1=2\; , \qquad \nu_1=2+{1 \over 1+2\alpha}\; , \qquad
\lambda_2={2 \alpha \over 2 \alpha+1}\; , \qquad \nu_2={-1 \over 2
\alpha+1}\; , 
\ee
\be
\beta= {\alpha(\alpha+1) \over 2\alpha+1} \; ,  
\qquad \gamma=-1-{1 \over 2\alpha+1} \; .
\ee
Here $\Phi(\lambda,\nu,y)$ stands for the confluent hypergeometric
function. It is regular at $ y=0 $ and grows exponentially for $ y \to
\infty $. Expression (\ref{solfluct}) is degenerate for $\alpha=-{1 \over 2}$:
\be
\alpha=-{ 1 \over 2} \; , \qquad f(y)= {B \over 1 + {y \over 4} } \; .
\ee
According to (\ref{varsep}), it diverges when the  time  $ \tau $
reaches $ 1 $. 

Another degenerate case is $\alpha=-1$, 
\be
\alpha=-1 \; , \qquad f(y)= y \left[ \ln \left({B_1 \over y}\right) +1
\right] + B_2  \; .
\ee
Interestingly, 
$ Re(\alpha) \in [-1,-{1 \over 2}] $ are the only values of $ \alpha $
where the fluctuations are regular at $ y=0 $. But again the fluctuations
diverge for $ \tau \to 1 $.

\begin{figure}[t!]
\begin{center}
\epsfig{width=17cm,file=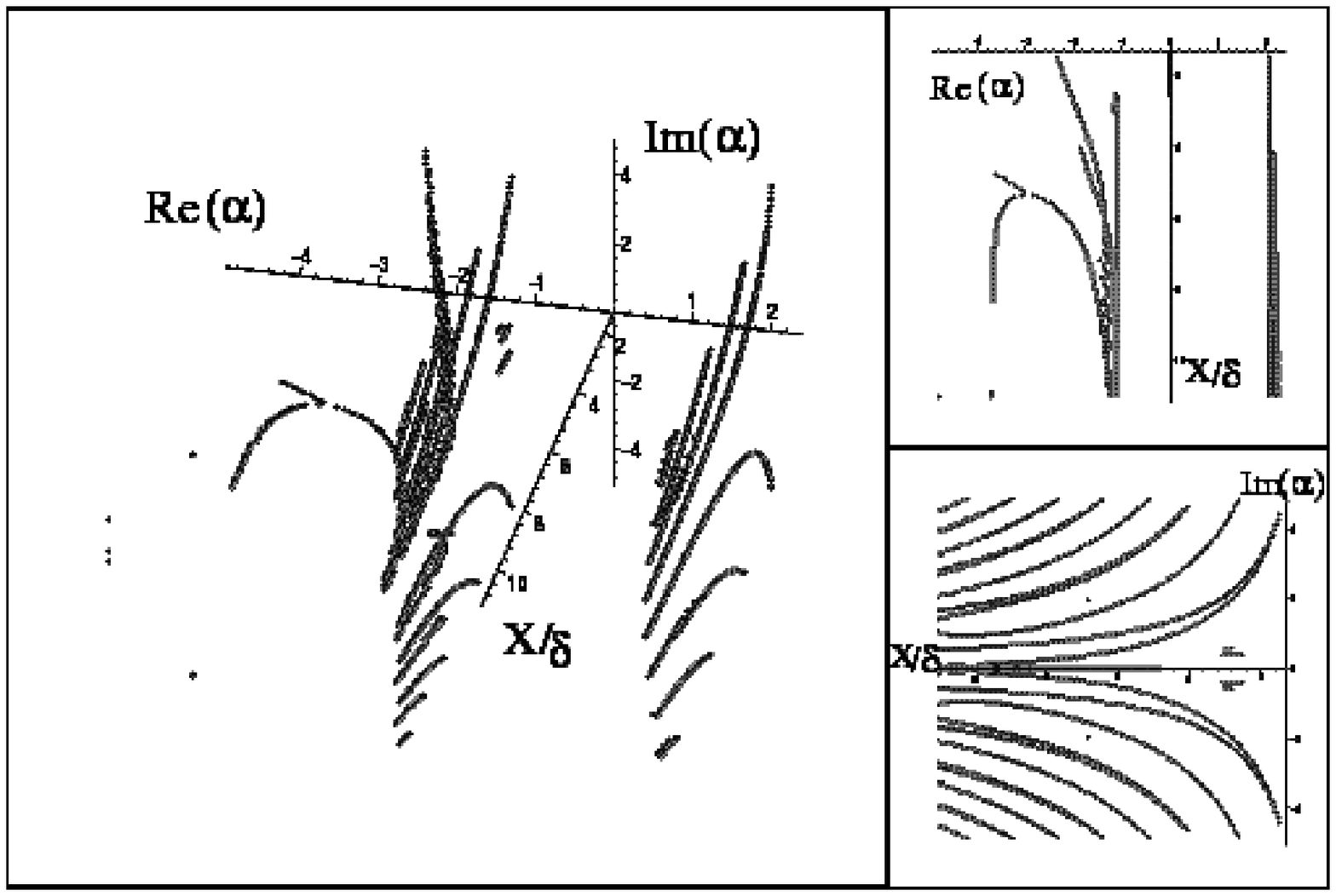}
\end{center}
\caption{\small Spectrum of modes around solution (\ref{solitsol}) for a 
continuous range of values of 
${R \over \delta}$. The real and imaginary parts of $\alpha$ are
plotted against 
${R \over \delta}$.}
\label{solitf}
\end{figure}

Since the fluctuations are ill-behaved for $ y \to \infty $  and often
at $ y = 0$, we
resort to the same method as in the singular isothermal sphere. We define
a large distance cutoff $X$ and a small distance cutoff $\delta$ for $y$ and we
set the fluctuation to zero on these walls. It should be noted that these
walls move with the soliton, and that they cannot be considered as physical
since the constant component of the background velocity field flows through 
the walls. However, those cutoffs allow us to search for fluctuations  as 
eigenmodes in the same way we did for isothermal spheres. These modes appear
for special values of $ \alpha $ defined here again by the cancellation of
a determinant:

\be
\Phi(\lambda_1,\mu_1,\beta \delta) X^\gamma \Phi(\lambda_2,\mu_2,\beta X) -
\Phi(\lambda_1,\mu_1,\beta X) \delta^\gamma \Phi(\lambda_2,\mu_2,\beta \delta)
\,=\,0
\ee
These cancellations must be investigated in the complex plane for $\alpha$ and
for various values of ${X \over \delta}$. The results
appear on fig. \ref{solitf}.
The main result is that modes with $ \hbox{Re}(\alpha)  <0 $ exist for
all values of 
$ {X \over \delta} $ investigated. These modes have a faster growth than the
background. Besides, $ {X \over \delta} \sim 4.5 $ is a special
value for the radial size of the system. Below this value all the unstable 
modes have the following time behaviour
$$ 
(1-\tau)^{\alpha_1} \cos[\alpha_2 \ln(1-\tau) + \phi]
$$
that is, growing oscillations.
Above this value  $ {X \over \delta}\sim 4.5  $ two modes appear that have a
pure power law behaviour in time. These modes have larger exponents than the
others, that is to say they grow faster. 
Moreover, just above this critical value, the exponent in the power law seems
to go to $-\infty$, which indicates a strong instability. 

We must keep in mind the fact that this is a partial analysis.
Indeed our fluctuations
are invariant along the $z$ symmetry axis. This means that they have an infinite
mass. If the size of the system along the $z$ symmetry axis was bounded and
fluctuations with $r$ and $z$ dependence computed, a stable region would
certainly appear for small enough sizes. We believe that the value
${X \over \delta} \sim 4.5$ is actually connected to the onset of a {\sl radial}
instability, while below this value the growth of the modes is due to the
axial infinite size of the fluctuations.

\section{ Conclusions}
\label{seven}
In the first part of this work we have described a general method to study
the dynamic stability of stationary configurations of
self-gravitating fluids and applied it to the isothermal 
spheres. The interest of this method is that it can be applied to any 
stationary field. Moreover, we have shown that it is perfectly able to
handle viscosity.  
We have applied this dynamical stability analysis to a stationary solution,
the isothermal sphere. In this case we have found a series of sizes
($X_{c_n} \sim 10.7^n,\; n=1,2,\ldots$) associated with the appearance
of instable modes. The  
size associated to the first instable mode matches the critical size 
found in the thermodynamic theory. Moreover, we have shown that the
values of these sizes are independent of the Reynolds number.

In the second part of this paper we studied exact dynamical solutions.
We have presented elements of a soliton theory for a
self-gravitating perfect isothermal fluid with axial symmetry. This
method provides new dynamic exact solutions. We have
analyzed the stability of one of these solutions, a collapsing filament,
with a method similar to the method used for the isothermal sphere. Here
again we have found that a critical radial size appears to define two regimes. 
Above
this size the system is unstable. Below, it is  weakly unstable and may
in fact be stable if the axial size of the system was bounded.

These two studies are complementary descriptions of the dynamics of the
self-gravitating fluid. One description deals with a stationary
background and the other 
with a dynamical one. Although the first description obeys spherical
symmetry and the 
second obeys axial symmetry, they are closely connected by their structure.
Indeed, the density field of eq.(\ref{solitsol})  is the sum of two terms:
$ \ln(2/\mu^2 r^2) $  (identical to the stationary spherical solution) and
a second term, $ \ln\left[  {\sqrt{C} \over 1 \pm \mu c_s t } \right] $,
 function of the soliton-time variable alone. In the
stationary case, the instability appears when a mode is growing;
in the dynamical case, it appears when a mode is growing faster (or
decaying slower) than the evolving background solution.
In this respect, the stability of the solution in both cases is
governed by 
Jeans-like instabilities whose emergence  depends only on the size of the 
system. We showed that the existence of a critical size does not only apply 
to stationary solutions but to dynamic solutions as well.

Finally, we would like to add that it is probably possible to give a more 
general formulation of the soliton-type equations of the system,
following on the results in ref.\cite{Gozt} in the one dimensional-case. For
example, the choice  
of the soliton variable in our work, eq.(\ref{solitvar}), is probably a 
particular case of a more general change
in the variables. In any case, the soliton methods provide  a powerful
approach to self-gravitating fluid dynamics which certainly deserves more
investigation.

\end{document}